\newif\ifconfver
 \confvertrue        

\ifconfver
	\documentclass{article}
	\usepackage{spconf}
	\title{MINIMUM SYMBOL ERROR RATE-BASED CONSTANT ENVELOPE PRECODING  FOR MULTIUSER Massive MISO Downlink}
	\name{Mingjie Shao$^{\dagger}$, Qiang Li$^{\star,\dagger}$, Wing-Kin Ma$^{\dagger}$ and Anthony Man-Cho So$^{\S}$
    \thanks{This work was supported in part by the Fundamental Research Funds for the Central Universities under Grant ZYGX2016J011.}
    }

	\address{
	\normalsize{
	$^{\dagger}$ Department of  Elec. Eng., The Chinese University of Hong Kong, Hong Kong SAR, China	} \\
	\normalsize{
	$^{\star}$ School of Info. \& Comm. Eng., University of Electronic Science and Technology of China, China}\\
	\normalsize{
	$^{\S}$ Department of Sys. Eng. \& Eng. Mgmt., The Chinese University of Hong Kong, Hong Kong SAR, China}\\
	\small {E-mail: $^{\dagger}$ \{mjshao, wkma\}@ee.cuhk.edu.hk, $^{\star}$  lq@uestc.edu.cn, $^{\S}$ manchoso@se.cuhk.edu.hk }
	}
\else
	\documentclass[11pt]{article}
	\usepackage{fullpage}
	\title{MINIMUM SYMBOL ERROR RATE-BASED CONSTANT ENVELOPE PRECODING  FOR MULTIUSER MASSIVE MISO DOWNLINK}
	
\fi
\usepackage{amsmath,graphicx}

\usepackage{calc,amsfonts,amssymb,bm,url,color,theorem,cite}
\usepackage{psfrag,subfigure,float}
\usepackage{algorithm}
\usepackage{algorithmic}
\usepackage{mathtools,lipsum,cuted,multicol}
\usepackage{filecontents}
\usepackage{epstopdf}
\usepackage{array}
\usepackage{caption}

\definecolor{orange}{RGB}{255,107,0}

\newtheorem{Fact}{Fact}

\theorembodyfont{\rmfamily}

\newtheorem{Remark}{Remark}

\newcommand\bs{\ensuremath{{\bm s}}}

\newcommand\bw{\ensuremath{{\bm w}}}

\newcommand\bh{\ensuremath{{\bm h}}}

\newcommand{\setD}{\mathcal{D}}
\newcommand{\setX}{\mathcal{X}}

\newcommand{\Rbb}{\mathbb{R}}
\newcommand{\Cbb}{\mathbb{C}}

\newcommand\bx{\ensuremath{{\bm x}}}

\newcommand\bH{\ensuremath{{\bm H}}}

\newcommand\bz{\ensuremath{{\bm z}}}

\newcommand\bbx{\ensuremath{\bar{\bm x}}}

\newcommand{\Rfrak}{\mathfrak{R}}
\newcommand{\Ifrak}{\mathfrak{I}}

\newcolumntype{M}[1]{>{\centering\arraybackslash}m{#1}}

\makeatletter
\def\bstctlcite{\@ifnextchar[{\@bstctlcite}{\@bstctlcite[@auxout]}}
\def\@bstctlcite[#1]#2{\@bsphack
  \@for\@citeb:=#2\do{%
    \edef\@citeb{\expandafter\@firstofone\@citeb}%
    \if@filesw\immediate\write\csname #1\endcsname{\string\citation{\@citeb}}\fi}%
  \@esphack}
\makeatother

\begin{document}
%
\bibliographystyle{IEEEtran}


\bstctlcite{IEEEexample:BSTcontrol}
\maketitle
\begin{abstract}

This paper considers the problem of constant envelope (CE) precoder designs for multiuser massive MISO downlink channels.
The use of CE signals allows one to employ low-cost radio frequency chains and thereby facilitates the implementation of massive MIMO.
However, the subsequent CE precoder designs are usually challenging owing to the non-convex CE constraints.
The existing CE precoder designs consider minimization of some measures on the distortion levels of the received symbols, and they usually aim at improving the symbol-error rate (SER) performances.
In this paper we formulate a minimum SER-based design for CE precoding.
The design formulation is non-convex and we propose two low-complexity first-order algorithms using gradient projection.
Curiously, our simulation results show that the proposed designs can achieve bit-error rate performance close to that of zero-forcing beamforming without CE signaling restrictions.

\end{abstract}
\begin{keywords}
massive MIMO, constant envelope, symbol error rate, non-convex projected gradient
\end{keywords}

\section{Introduction}
Massive MIMO, as one of the core physical-layer techniques for the fifth generation mobile system, can provide substantial spectral efficiency gains \cite{Rusek2013,Lu2014}. In order to fully harness the benefits of massive MIMO, high-performance radio-frequency (RF) chains with large linear dynamic range are desired. However, this would push up the hardware cost tremendously as the number of antennas scales  to hundreds or even more. To circumvent this difficulty, a signal processing-based constant envelope (CE) precoding solution was proposed \cite{Mohammed2012}, where the amplitudes
of the transmit signals are fixed and only the phases are changed from symbol to symbol.
Due to the low peak-to-average power ratio (PAPR) of CE signals, one can use very cheap RF chains (with limited dynamic range) to amplify  the signals without incurring much distortion.

While CE precoding can be easily implemented with phase shifters, the design of CE precoder itself is a challenging task, owing to the non-convex nature of the CE constraints. In light of this, a great deal of efforts have been devoted to CE precoder designs under various constellations and user settings.  In \cite{Mohammed2012} and \cite{Pan2014}, a full study for the feasibility of CE precoding in single-user MISO channels was done and an optimal phase recovery algorithm for CE precoding was proposed in \cite{Pan2014}.
Subsequently, the concept of CE precoding was extended for multiuser MISO \cite{Mohammed2013,Chen2014,Mohammed2013a,Amadori2017,Liu2017}  and point-to-point MIMO \cite{Zhang2017}. In particular, the works \cite{Mohammed2013,Chen2014} and \cite{Mohammed2013a} studied the multiuser interference (MUI) power minimization by CE precoding under frequency-flat channel and frequency-selective channel, respectively.
In addition to MUI reduction, the concept of constructive interference using CE precoding is investigated in \cite{Amadori2017} for PSK modulations; cross-entropy  and convex relaxation methods are proposed to handle the CE problem.
The joint optimization of CE precoding and receive beamforming is considered for a single-user MIMO system in \cite{Zhang2017}.
Except for the single-user MISO case, the vast majority of the aforementioned works adopt some symbol distortion measures as their guidelines for CE precoder designs.
In those studies it is usually anticipated that minimizing those symbol distortion measures should lead to improved symbol-error rate (SER) performance.

In this work we consider a minimum SER-based design for CE precoding in multiuser massive MISO downlink systems. Our minimum SER-based design formulation follows that of our very recent work for one-bit MIMO precoding \cite{mjshao2018} --- which, by nature, is even more difficult than CE precoding and we handled it via a sophisticated alternating minimization algorithm.
For self-containedness we will describe the formulation in Section \ref{sec:format}.
In the present paper, we will propose a simple non-convex projected gradient (PG) algorithm for the CE precoder design problem.
To further speed up the convergence of the non-convex PG, we also consider the FISTA-type acceleration~\cite{Beck2009}.
Our simulation results show that the bit-error rate (BER) performance of our CE design for $16$-QAM and $64$-QAM is about $1-3$dB away from that of zero-forcing beamforming without stringent CE constraints; and the acceleration is helpful in reducing the running times.

\section{System Model and Problem Formulation}
\label{sec:format}
\subsection{System Model}
Consider a multiuser MISO downlink scenario where the base station (BS) equipped with $N$ antennas unicasts $K$ information symbol streams  to $K$ single-antenna users. Assuming block fading channel and perfect channel state information (CSI) at the BS, the received signal at user $i$ during the $t$th time slot of fading block $\ell$ is given by
\begin{equation} \label{model}
\begin{aligned}
y_{i,t}=\bh_{i,\ell}^T \bx_t + n_{i,t} ,&~~i=1,\ldots, K,\\
&~~t\in ((\ell-1)T, ~\ell T]
\end{aligned}
\end{equation}
for $\ell =1,2,\ldots$, where $T$ is the block length; $\bh_{i,\ell}^T$ is the downlink channel for user $i$ during fading block $\ell$; $n_{i,t}$ is additive white Gaussian noise with mean $0$ and variance $\sigma_n^2$; $\bm x_t \in \mathbb{C}^N$ is the transmit signal at the BS. As mentioned before, CE precoding is employed for facilitating massive MIMO implementation. In CE precoding, $\bx_t$ conforms to the following constraint:
 \[
\bm x_t \in  \setX \triangleq \left\{\bx\in \Cbb^{N}~|~|x_i|^2=P/N,~~i=1,\ldots,N\right\},
 \]
where $P$ is the total transmission power at the BS; here we assume that each antenna has equal transmit power. In the following derivations, we focus on one fading block and drop the fading block index $\ell$ for notational simplicity. Similarly, the time slot index $t$ will be suppressed when there is no ambiguity.

The basic idea of CE precoding is to design the transmit signal $\bm x$ so that the noise-free receive signal at the $i$th user, given by $\bm h_i^T \bm x$, is close to user $i$'s desired symbol $s_i$, i.e.,
\begin{equation}\label{eq:CE_symbol}
    \bh_i^T \bx \approx d \cdot s_{i},~\forall~i=1,\ldots, K.
\end{equation}
Herein, $d\geq 0$ is a symbol shaping gain factor, which depends on $\bm x$ and needs to be optimized; $s_{i}$'s are drawn from a QAM constellation $\mathcal{S}$, viz.
\[
    \mathcal{S}=\left\{ s_{R}+j s_{I}~|~s_{R},s_{I}\in \{ \pm 1,\pm 3,\ldots, \pm (2L-1) \} \right\}
\]
with $L$ being the order of the QAM constellation. From Eqns.~\eqref{model} and \eqref{eq:CE_symbol}, the symbol detection at users is easily performed as
\[
    \hat{s}_{i}=\text{dec}(y_{i}/ d),
\]
where $\text{dec}(\cdot)$ is a decision function, mapping $y_{i}/ d$ to the nearest constellation symbol in $\cal S$. For simplicity, we assume that all the users know $d$  {\it a priori}, say, by estimating $d$ from the training symbols before data block transmission or by broadcasting $d$ from the BS to users via a side channel.

\subsection{Problem Statement}
Our goal is to minimize the worst SER among all the users. To proceed, let us first characterize user $i$'s SER as
$$ \text{SER}_{i}=\text{Pr} (\hat{s}_{i} \neq s_{i}| s_{i}).$$
Notice that
\begin{equation}\label{SERUP}
\begin{split}
\text{SER}_{i} \leq \text{SER}_{i}^{R} + \text{SER}_{i}^{I}\leq 2\max
\{ \text{SER}_{i}^{R}, \text{SER}_{i}^{I} \},
\end{split}
\end{equation}
where $\text{SER}_{i}^{R} \triangleq \text{Pr}(\Rfrak\{\hat{s}_{i}\}\neq \Rfrak\{s_{i}\}|s_{i})$ and $\text{SER}_{i}^{I} \triangleq \text{Pr}(\Ifrak\{\hat{s}_{i}\}\neq \Ifrak\{s_{i}\}| s_{i})$ correspond to the
error probabilities of the in-phase  and the quadrature components, respectively. Moreover, a straightforward calculation gives
\begin{equation}\label{RIUB}
  \begin{split}
    \text{SER}_{i}^{R}&\leq 2 Q\left(  \frac{d-|\Rfrak\{\bh_i^T \bx\}-d\Rfrak\{s_{i}\} | }{\sigma_n/ \sqrt{2}} \right)  \triangleq M_{i}^R ,\\
    \text{SER}_{i}^{I} &\leq  2 Q\left( \frac{d-|\Ifrak\{\bh_i^T \bx\}-d\Ifrak\{s_{i}\} | }{\sigma_n/ \sqrt{2}} \right) \triangleq M_{i}^I,
  \end{split}
\end{equation}
where $Q(x)=\int_{x}^{\infty} \frac{1}{\sqrt{2\pi}} e^{-z^2/ 2} dz$, $\Rfrak\{\cdot\}$ and $\Ifrak\{\cdot\}$ denote the real part and the imaginary part, respectively. Using~\eqref{SERUP} and \eqref{RIUB}, user $i$'s SER can be upper bounded by
\begin{equation}\label{eq:upper_bound}
  \text{SER}_{i}\leq 2 \max\{M_{i}^R, M_{i}^I \},~\forall~i=1,\ldots, K.
\end{equation}

\par From \eqref{eq:upper_bound}, we consider the following worst SER-based CE precoding problem:
\begin{equation}\label{UB}
  \begin{split}
  \min_{ \bx,d}&~~\max_{i=1,\ldots, K}~\max \{ M_{i}^R, ~M_{i}^I \}\\
  \text{s.t.} &~~ \bx \in \setX,~~ d\geq 0,
  \end{split}{}
\end{equation}
which can be equivalently written as the following concise form by exploiting the monotonicity of the $Q$ function:
\begin{equation} \label{Form1}
  \begin{split}
    \min_{\bbx,d }&~~\max_{i} |\bar{\bh}_i^T \bar{\bx}-d\bar{s}_{i}|-d,\\
    \text{s.t.}&~~ \bar{\bx} \in \setX_{\Rfrak},~~ d\geq 0,
      \end{split}
\end{equation}
where
\begin{align*}
\setX_{\Rfrak} & \triangleq \{\bx\in \Rbb^{2N}|~|x_j|^2\!+\!|x_{j+N}|^2={P}/{N},~j=1,\ldots,N \}, \\
 \bar{\bm x}& = [\Rfrak\{\bx\}^T,~ \Ifrak\{\bx\}^T]^T, ~~  \bar{\bs} = [\Rfrak\{\bs\}^T,~\Ifrak\{\bs\}^T]^T, \\
 \bH & =[\bh_1, \bh_2,\ldots,\bh_K]^T, \\
 \bar{\bH} & =[\bar{\bh}_1, \ldots, \bar{\bh}_{2K}]^T=\begin{bmatrix}
               \Rfrak\{\bH\} & -\Ifrak\{\bH\} \\
               \Ifrak\{\bH\} & \Rfrak\{\bH\}\end{bmatrix}.
\end{align*}

Problem~\eqref{Form1} is derived under a single time slot.
The same formulation can be derived for the case of one fading block. Letting $\{\bar{\bm x}_t\}_{t=1}^T$ be the equivalent real-valued transmit signals within one fading block and following the derivations above,  the  worst SER-based CE precoding for block transmission can be shown to be
\begin{equation} \label{final}
  \begin{split}
    \min_{\{\bbx_t\}_{t=1}^T,d }&~~\max_{\substack{i=1,\ldots, K,\\t=1,\ldots, T}} |\bar{\bh}_i^T \bar{\bx}_t-d\bar{s}_{i,t}|-d\\
    \text{s.t.}&~~ \bar{\bx}_t \in \setX_{\Rfrak},~~ d\geq 0,~~t=1,\ldots, T.
      \end{split}
\end{equation}
\begin{Remark}
  Problem \eqref{final} is physically sound. Specifically, the objective consists of two parts --- One is the distortions of the noise-free received symbols; the other is the symbol shaping gaining factor $d$. Intuitively, small distortion and large symbol shaping gain factor would make the CE design robust against noise. By jointly optimizing $\{\bar{\bm x}_{t}\}_{t=1}^T$ and $d$, problem~\eqref{final} can provide a better balance between the two metrics, as compared with  most existing CE literatures where $d$ is fixed \cite{Mohammed2012,Mohammed2013,Chen2014}.
\end{Remark}


\par The resulting CE precoding problem~\eqref{final} is a  non-smooth non-convex problem.
In the following section, we will develop a non-convex projected gradient-based algorithm for problem \eqref{final}.

\section{Proposed CE precoding algorithm}

Our development consists of two steps: First, we tackle the non-smooth objective in~\eqref{final} by a smooth approximation. Second, a non-convex projected gradient-based method is applied to handle the smoothed problem.

\subsection{Smooth Approximation of \eqref{final}}
We apply the well-known log-sum-exp inequality to smoothen the objective. Specifically, note the following fact~\cite{CVX}:
\begin{Fact}
Given $a_1\ldots,a_K\in \mathbb{R}$,  it holds for any $\sigma> 0$ that
\begin{equation}\label{logsumexp}
 \max_{i=1,\ldots,K} a_i\leq \sigma\log \sum_{i=1}^K e^{\frac{a_i}{\sigma}}\leq \max_{i=1,\ldots,K} a_i +\sigma \log K.
\end{equation}
Moreover, the inequalities become tight as $\sigma\rightarrow 0$.
\end{Fact}
Utilizing \eqref{logsumexp}, the CE precoding problem \eqref{final} is smoothly approximated as
\begin{equation} \label{smooth}
  \begin{aligned}
    \min_{\bbx_t,d }&~~~f(d,\{\bar{\bm x}_t\}_{t=1}^T) \\
    \text{s.t.}&~~~ \bar{\bx}_t \in \setX_{\Rfrak},~~ d\geq 0,~~t=1,\ldots, T.
  \end{aligned}
\end{equation}
where $$f(d,\{\bar{\bm x}_t\}_{t=1}^T) \triangleq \!\sigma \! \log \sum_{i,t}\left[ e^{\frac{\bar{\bh}_i^T \bar{\bx}_t-d\bar{s}_{i, t}-d}{\sigma}}+e^{\frac{-\bar{\bh}_i^T \bar{\bx}_t+d\bar{s}_{i, t}-d}{\sigma}} \right].$$

\subsection{Non-convex Gradient Projection Algorithm for \eqref{smooth}}
For ease of exposition, we define
$$\setD \triangleq \{(d, \bar{\bm x}_1,\ldots, \bar{\bm x}_T)~|~d\geq 0, ~~\bar{\bx}_t \in \setX_{\Rfrak},~t=1,\ldots, T\}$$
to be the feasible set of problem~\eqref{smooth}. Let $\bm z \triangleq (d, \bar{\bm x}_1,\ldots, \bar{\bm x}_T)$.
The proposed gradient projection method recursively updates $\bm z$ according to the following equation
\begin{equation} \label{eq:gp}
  \bz^{l+1}\!=\Pi_{\setD} \left(\bz^l\!-\!\gamma_l\nabla f(\bz^l)\right),
  \end{equation}
where the superscript $l$ denotes the iteration number, and $\gamma_l>0$ is the stepsize, which can be determined by backtracking line search~\cite{Beck2009}; $\Pi_{\cal D}(\bm x)$ denotes the projection of $\bm x$ onto the  set $\cal D$. While $\cal D$ is non-convex, the projection can be easily computed in closed form. Specifically, let $$\tilde{\bm z}^l= \left(\tilde{d}^l, \tilde{\bm x}_1^l,\ldots, \tilde{\bm x}_T^l \right) =\bz^l\!-\!\gamma_l\nabla f(\bz^l) .$$ Then,
\begin{equation}\label{eq:PG_update}
 \begin{cases}
    d^{l+1}   = \max\{0, \tilde{d}^l \}, \\
    [\bar{\bm x}_t^{l+1}]_{j}   = \sqrt{\frac{P}{N}}\frac{[\tilde{\bm x}_t^l]_j}{\sqrt{|[\tilde{\bm x}_t^l]_j|^2+|[\tilde{\bm x}_t^l]_{j+N}|^2}},~j=1,\ldots, N, ~\forall~t,\\
    ~[\bar{\bm x}_t^{l+1}]_{j} \!\!  = \!\! \sqrt{\frac{P}{N}}\frac{[\tilde{\bm x}_t^l]_{j}}{\sqrt{|[\tilde{\bm x}_t^l]_{j-N}|^2+|[\tilde{\bm x}_t^l]_{j}|^2}},j\!=\!N\!+1,\ldots, 2N,\forall~t,\\
 \end{cases}
\end{equation}
where $[\bm x]_j$ denotes the $j$th element of $\bm x$. The detailed algorithm  is summarized in Algorithm~\ref{Armijo}.

By adapting the proof in \cite{Tranter2017}, it can be shown that every limit point, denoted as $\bm z^\star$, generated by Algorithm~\ref{Armijo} is a stationary point of problem~\eqref{smooth}. We omit the proof due to the page limit and will present it in the journal version.

%

\begin{algorithm}[htb!]
\caption{: Projected Gradient for CE Precoding}
\begin{algorithmic}[1]
\STATE Initialize  $\bz^{0}=(\bar{\bx}^{0},d^{0})$, $\sigma$ and $l=0$
\REPEAT
\STATE Calculate $\nabla f(\bm z^l)$ as
\begin{equation*}
  \begin{split}
\frac{\partial f(\bm z^l)}{\partial \bbx_t}&=\frac{\sum\limits_{i}\left(W_{i,t}^P- W_{i,t}^N\right)\bar{\bh}_i}{\sum\limits_{i,t}\left(W_{i,t}^P+W_{i,t}^N \right)},~~t=1,\ldots, T,\\
\frac{\partial f(\bm z^l)}{\partial d}&=\frac{\sum\limits_{i,t}\left(-W_{i,t}^P(\bar{s}_{i,t}+1)+W_{i,t}^N(\bar{s}_{i,t}-1) \right)}{\sum\limits_{i,t}\left(W_{i,t}^P+W_{i,t}^N \right)},
  \end{split}
\end{equation*}
where $W_{i,t}^P=\exp\left({\frac{\bar{\bh}_i^T \bar{\bx}_t^l-d^l\bar{s}_{i, t}-d^l}{\sigma}}\right)$ and $W_{i,t}^N=\exp\left({\frac{-\bar{\bh}_i^T \bar{\bx}_t^l+d^l\bar{s}_{i, t}-d^l}{\sigma}}\right)$;
\STATE Update $d^{l+1}$ and $\{\bar{\bm x}_t^{l+1}\}_{t=1}^{T}$ according to Eqn.~\eqref{eq:gp} and Eqn. \eqref{eq:PG_update};
 \STATE $l=l+1$;
\UNTIL {some stopping criterion is satisfied.}
\end{algorithmic}\label{Armijo}
\end{algorithm}
~\\[-3.5em]

\subsection{Fast Non-convex Gradient Projection for \eqref{smooth}}
In this section, we introduce an acceleration scheme to further speed up the convergence of Algorithm~\ref{Armijo}. The idea is similar to the FISTA algorithm~\cite{Beck2009} or Nesterov's accelerated gradient method~\cite{IntroCVX}, though FISTA was originally developed for convex problems. Specifically, we modify Eqn.~\eqref{eq:gp} as
\begin{equation*}
   \bz^{l+1}\!=\Pi_{\setD} \left(\bw^l\!-\!\gamma_l\nabla f(\bw^l)\right),
\end{equation*}
where $\bm w^l$ is an extrapolated point of $\bm z^l$ and $\bm z^{l-1}$, i.e.,
\begin{equation*}
\begin{aligned}
 \bw^{l}&=\bz^{l}+\frac{\beta_{l}-1}{\beta_{l+1}}(\bz^{l}-\bz^{l-1}),\\
 \beta_{l+1}&=\frac{1+\sqrt{1+4 \beta_{l}^2}}{2},
 \end{aligned}
 \end{equation*}
with $\bz^{0}=\bz^{-1}$ and $\beta_0 = 1$.

It has been shown in~\cite{Beck2009} that  this acceleration can improve the convergence  rate of the projected gradient method from $\mathcal{O}(1/l)$ to $\mathcal{O}(1/l^2)$ for convex problems. While our considered problem is non-convex, our numerical experience suggests that this acceleration scheme is still very effective, as illustrated in the ensuing section.

\section{Simulation Results}
\label{sec:sim}

In this section, we evaluate the performance of our proposed algorithms by Monte-Carlo simulations. Three schemes are compared, namely, zero-forcing (ZF) without CE constraints, named ``ZF''; ZF with naive projection onto the CE set, named ``CE ZF'';  and the total MUI power minimization algorithm~\cite{Mohammed2013}, named ``MUImin''. We will use ``PG'' and ``FPG'' to represent Algorithm~\ref{Armijo} and its accelerated version.
BER is used as the performance metric.

The simulation setting is as follows: A block Rayleigh fading channel is assumed with transmission block length $T=10$. The total transmit power is $P=1$. The elements of channel $\{\bh_i\}_{i=1}^K$ are i.i.d. generated according to $\mathcal{CN}(0,1)$.  Both $16$-QAM and $64$-QAM modulations are considered. For both Algorithm \ref{Armijo} and its accelerated variant, the smoothing parameter is set to $\sigma=0.05$, and the algorithms stop when the  improvement of successive iterations is less than $10^{-4}$, or the maximum number of iterations $5,000$ is reached. All the results were averaged over $10^4$ independent channel trials.
\begin{figure}[htb!]
\centering
\includegraphics[width=0.92\linewidth]{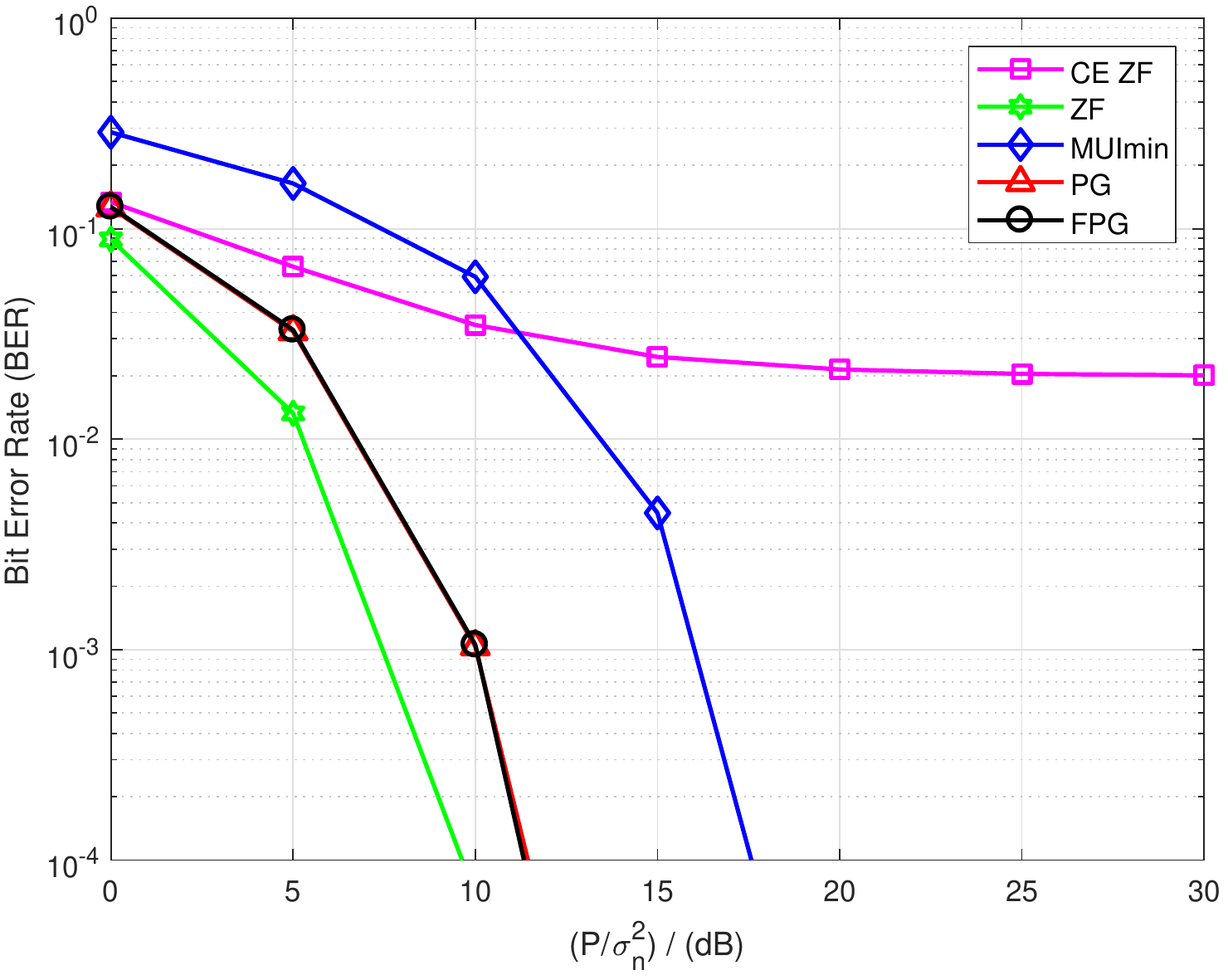}
\caption{Average BER performance versus $P/\sigma_n^2$; $16$-QAM.}\label{16sim}
\end{figure}

\begin{figure}[htb!]
\centering
\includegraphics[width=0.92\linewidth]{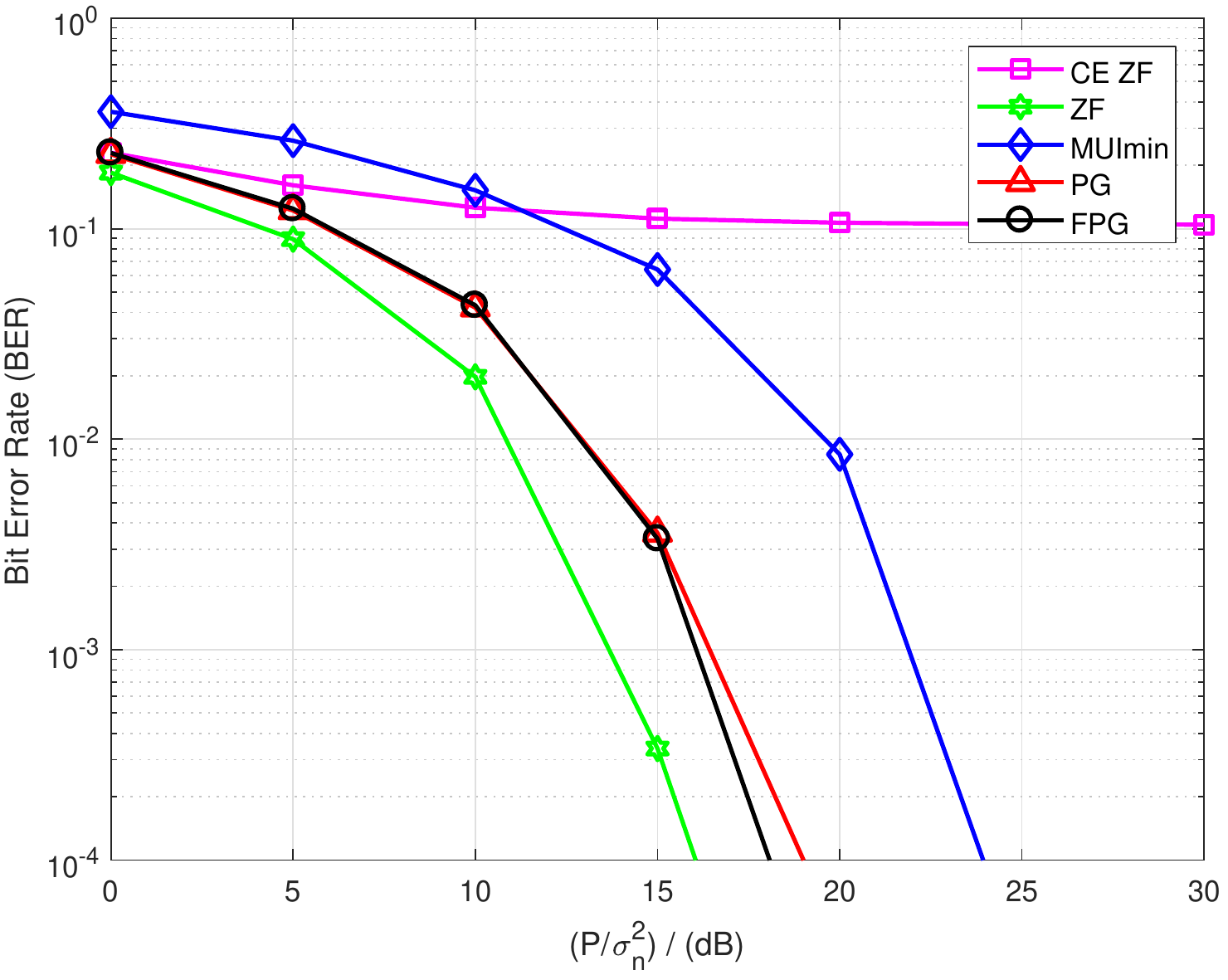}
\caption{Average BER performance versus $P/\sigma_n^2$; $64$-QAM.}\label{64QAM}
\end{figure}
In Fig. \ref{16sim}, we compare the BER performance under $16$-QAM modulation scheme; the number of transmit antennas is $N=128$ and the number of users is $K=16$.
It is seen that both PG and FPG achieve much better BER performance than CE ZF and MUImin.  Notice that the performance gap between ZF (without CE constraints) and FPG is only about $2.5$dB at BER$=10^{-3}$.
Fig. \ref{64QAM} shows the simulation result under the same setting as Fig. \ref{16sim}, except that $64$-QAM is used. The figure shows similar performance behaviors.

While PG and FPG have similar BER performances, FPG is much more computationally efficient than PG. To verify this, we compared the runtime of PG and FPG under different number of antennas with fixed number of users and modulation scheme. The results are shown in Table \ref{my-label}. As seen, FPG can achieve significant  runtime reduction, especially for large problem sizes.
\\[-1.3em]
\begin{table}[H]
\centering
\captionsetup{justification=centering}
\caption{Average runtime (in Sec.) for each transmission block ($K=16$, 64-QAM)\\[-0.4em]}\label{my-label}
\renewcommand{\arraystretch}{1.2}
\resizebox{\linewidth}{!}{%
\begin{tabular}{M{10mm}|M{15mm} M{15mm} M{15mm} M{15mm} }
\hline
$N$ & 50 & 100 & 150 & 200 \\ \hline\hline
PG & 0.531 & 0.701 & 1.053 & 1.29 \\
FPG  & 0.415 & 0.436 & 0.545 & 0.607 \\ \hline
\end{tabular}
}
\end{table}
\section{Conclusion}
In this paper, we have considered the CE precoder design for multiuser massive MISO downlink channels. A worst SER-based CE formulation is employed. By exploiting the problem structure of the CE precoding, a simple and efficient non-convex projected gradient algorithm and its accelerated variant were derived. Simulation results showed that our proposed algorithms can achieve better BER performance than the existing CE precoder designs.

\vfill\pagebreak
\newpage

\bibliographystyle{IEEEtran}
\bibliography{refs}
\nocite{*}

\end{document}